\documentclass[12pt]{article}
\title{Classification of transformations of  probabilities for preparation procedures:
trigonometric and hyperbolic behaviours.}

\author{Andrei Khrennikov\\
Department of Mathematics, Statistics and Computer Sciences\\
University of V\"axj\"o, S-35195, Sweden\\
Email:Andrei.Khrennikov@msi.vxu.se}

\begin{document}
\maketitle

\begin{abstract}  We provide frequency probabilistic analysis of
perturbations of  physical  systems by preparation procedures.
We obtained the classification of possible probabilistic
transformations connecting
input and output probabilities that can appear in physical experiments.
We found that  so called quantum probabilistic
rule is just one of possible rules. Besides the well known trigonometric
transformation (for example, for the polarization of light), there exist
the hyperbolic transformation of probabilities. In fact, 
`hyperbolic polarization'  have laready been observed 
in experiments with elementary particles. However, it was not interpreted 
in such a way. The situation is more complex with the hyperbolic interference
of alternatives.

\end{abstract}

We obtained the classification of possible probabilistic
rules that could connect input and output probabilities for preparation 
procedures (for micro as well as for macro systems), see [1] on the general 
theory of preparation/measurement procedures.
We found that the standard
trigonometric probabilistic transformation (so called `quantum probabilistic
rule') is just one of possible rules. Our analysis implies that, besides the 
trigonometric transformations of probability which appear in the quantum formalism,
there exist hyperbolic transformations of probability. Moreover, such transformations
connect input and output probabilities  in the well known experiments 
with elementary particles: trigonometric `polarization'
in some directions and hyperbolic `polarization' in other directions.

In fact, in this paper we follow to so called contextualist approach to physical probabilities,
see, for example, [1], [2] for the details. Probability distributions of physical
variables are determined not only by objective properties of physical systems,
but also by the whole context of a preparation/measurement procedure, see N. Bohr [3].
We point out that the contextualist framework is characterized by 
a large diversity of viewpoints. In particular, there are various viewpoints
on the origin of quantum stochasticity.  It might be that quantum stochasticity
can be reduced to classical stochasticity. It might be not. In fact, in the
present paper we demonstrate that quantum stochasticity can be reduced to
classical stochasticity (at least simulated for macro systems).

This paper is closely related to the paper [4] (which was also presented as a plenary lecture at International
Conference "Foundations of Probability and Physics", Vaxjo, Sweden-2000). We hope 
that this present text gives clearer presentation of our ideas.

\section{Classification of probabilistic transformations in Nature}

We shall use so called frequency approach 
to probability which was developed by R. von Mises [5], see also [6]. In 
that approach the probability is defined as the limit of relative frequencies. 
\footnote{Such an approach is quite natural for physicists, see, for example, A. Peres in [1].
However, it was strongly criticized in mathematical literature, see, for example, [6],
because there were some problems with the definition of randomness. We would not 
use the notion of randomness in our considerations. Therefore all this critique has no
relation to our considerations. The only thing that we shall use is that the probability
is nothing than the limit-value of the  corresponding relative frequency. 
The frequency approach to probability is very useful for probabilistic analysis
of physical phenomena. In such a way we can obtain some results that it would be
impossible to obtain in the conventional approach to probability that is based on the
abstract Kolmogorov axiomatics, [7].}

Let $S$ be an ensemble of physical systems.  Let $a (=a_1, a_2)$ be a dichotomic
physical variable 
which can be measured for elements of the ensemble $S.$
Probabilities of values $a_i, i=1,2,$ are defined as

${\bf p}_i = \lim_{N\to \infty} {\bf p}_i(N),\; \; {\bf p}_i(N)= \frac{n_i}{N} \;.$

Here $ N = \vert S\vert  $ and $n_i= \vert \{ s \in S: a=a_i\}\vert,$ where the symbol
$\vert O\vert$ denotes the number of elements in the ensemble $O$.

Let ${\cal E}$  be some preparation procedure; see [1] (see also P. Dirac [8]: ``In practice the conditions
could be imposed by a suitable preparation of the system, consisting perhaps in passing it through various 
kinds of sorting apparatus, such as slits and polarimeters, the system being left undisturbed after
the preparation.''). Suppose that by
applying ${\cal E}$ to the ensemble $S$ we produce a new ensemble $S^\prime.$
It is assumed that  ${\cal E}$ does not change the size of population: 

$N= \vert S\vert = \vert S^\prime\vert.$ 

All our considerations
are based on the simple remark  that a
preparation procedure may change the  probability distribution of values of $a.$ 
Probabilities of $a_i, i=1,2,$ after the preparation procedure ${\cal E}$ (output probabilities)
are defined as 

${\bf p}_i^\prime = \lim_{N\to \infty}{\bf p}_i^\prime(N) ,\; \; {\bf p}_i^\prime(N)= \frac{n_i^\prime}{N},\;
i=1,2,$

where  $n_i^\prime = \vert \{ s \in S^\prime: a=a_i\}\vert.$
We make the following trivial calculations:

${\bf p}_i^\prime(N)  = \frac{n_i^\prime}{N}= 
\frac{n_i}{N} + \delta_i(N) .$

Here 

$ \delta_i(N)= \frac{n_i^\prime - n_i}{N}= {\bf p}_i(N) \lambda_i (N),$

where 

$ \lambda_i (N)= \frac{n_i^\prime - n_i}{n_i}.$

Finally, we get
\begin{equation}
\label{LL}
{\bf p}_i^\prime(N)  = {\bf p}_i(N)  (1 + \lambda_i (N)).
\end{equation}

The coefficient $\lambda_i (N)$ gives the statistical deviation for  the 
distribution of
$a$ induced by the preparation procedure ${\cal E}. $ We note that there exist limits 

$ \lambda_i  = \lim_{N\to \infty} \lambda_i (N).$

This is a consequence of the existence of limits for relative frequencies 
${\bf p}_i(N)$ and ${\bf p}_i^\prime(N) .$\footnote{In fact, we (as always in modern 
physics) assume that the preparation procedure ${\cal E}$ is statistically regular:
all relative frequencies stabilize when $N \to \infty,$ compare with [6]
where we considered the possibility that the law of the statistical stabilization
(the law of large numbers) might be violated for some physical variables.} 
Thus by taking the limit when $N\to \infty$ in (\ref{LL}) we get the following relation 
between the input  and output probabilities:
\begin{equation}
\label{LL1}
{\bf p}_i ^\prime = {\bf p}_i  (1 + \lambda_i ).
\end{equation}
This is the general probabilistic transformation which could be produced by 
natural  (or even social) statistical phenomena.
We remark that the
coefficients $\lambda_1$ and $\lambda_2$ are not independent. There is the 
normalization condition:

$ 1= {\bf p}_1^\prime + {\bf p}_2^\prime= {\bf p}_1 + {\bf p}_2 + \lambda_1 {\bf p}_1 +
\lambda_2 {\bf p}_2.$

We obtained a kind of orthogonality relation:
\begin{equation}
\label{OR}
\lambda_1 {\bf p}_1 + \lambda_2 {\bf p}_2 =0 .
\end{equation}
Magnitudes of  the coefficients  $ \lambda_i$ will play the crucial role in our analysis.

{\bf (T)} Let $\vert \lambda_i \vert \leq 1, i=1,2.$  This means that the relative magnitude of perturbations of frequencies is not so large. Here 
\begin{equation}
\label{LL2}
\vert n_i^\prime - n_i \vert \leq n_i, N\to \infty.
\end{equation}

Thus perturbations induced by the ${\cal E}$ may change statistics strongly, but not 
crucially, compare with the case {\bf (HT).}  The coefficients $\lambda_i$ can be represented in the form

$\lambda_1 = \cos \theta_1,\; \; \; \lambda_2 = \cos \theta_2,$

where $\theta_1$ and $\theta_2$ are some `phases.' 
Orthogonality relation (\ref{OR}) implies that  phases 
$\theta_1$ and $\theta_2$ are not independent:
\begin{equation}
\label{ORT}
\cos \theta_1 {\bf p}_1 + \cos \theta_2 {\bf p}_2 =0 .
\end{equation}
Thus
\begin{equation}
\label{ORT1}
\frac{\cos \theta_2}{\cos \theta_1}= - \frac{{\bf p}_1}{{\bf p}_2} .
\end{equation}

Thus if magnitude of the statistical deviation induced  by a preparation 
procedure  is not so large, see (\ref{LL2}), then we obtain the following
rule for the transformation between input and output probabilities:
\begin{equation}
\label{T1}
{\bf p}_1^\prime = {\bf p}_1 (1+ \cos \theta_1)= 2 {\bf p}_1 \cos^2 \frac{\theta_1}{2} \;,
\end{equation}
\begin{equation}
\label{T2}
{\bf p}_2^\prime = {\bf p}_2 (1+ \cos \theta_2)= 2 {\bf p}_2 \cos^2 \frac{\theta_2}{2} \;.
\end{equation}
We call transformation  (\ref{T1}), (\ref{T2})  the {\it trigonometric probabilistic rule.}
In particular, if  $ {\bf p}_1={\bf p}_2= 1/2,$ we get 
(on the basis of (\ref{T1}), (\ref{T2}) and (\ref{ORT}))   
the  probabilistic
rule for {\it light  polarization}:

${\bf p}_1^\prime= \cos^2 \;\alpha, \; \; {\bf p}_2^\prime=\sin^2 \;\alpha,$ 

where $2 \alpha = \theta_1= \theta_2+\pi.$
We now consider an important particular case of  $T$-probabilistic behaviour.

{\bf (C)} Let a preparation procedure ${\cal E}$ produce negligibly
small statistical deviation:

$ \lim_{N \to \infty} 
\lambda_i(N) = \lim_{N \to \infty} 
\frac{n_i^\prime - n_i}{n_i}=0, i=1,2.$

In such a case we have $ {\bf p}_i ^\prime ={\bf p}_i, i =1,2.$ 
This is so called classical probabilistic behaviour (compare with P. Dirac [8], p. 11).

$T$-probabilistic behaviour (and, in  particular, classical and quantum behaviours) 
has been already  observed in various physical experiments. We are going to consider new probabilistic
behaviours which have not been yet observed in physical experiments. However, in our 
frequency considerations  those new probabilistic behaviours  are not less natural than $T$-behaviour.

We remark that the coefficients $\lambda_1$ and $\lambda_2$ have opposite signs. We can assume
that $\lambda_1\geq 0$ and $\lambda_2\leq 0.$ 
As $p_2^\prime \geq 0,$ we get that the coefficient $\lambda_2$ must always belong
to the interval $[-1,0].$
Thus this coefficient can be always
represented in the form:

$
\lambda_2 =\cos \; \theta_2.
$

On the other hand, $\lambda_1$ can be less as well as larger than 1. In the first case
we can represent it as $\lambda_1= \cos \; \theta_1;$ in the second case 
$\lambda_1= \cosh \; \; \theta_1.$ The first case has been already studied, see ${\bf (T)}.$
We now study the second case.

{\bf (HT)} Let $\lambda_1  > 1$ and $-1 \leq\lambda_2 \leq 0.$  
This means that the statistical deviation for $a=a_1$ is sufficiently large:

$\vert n_1^\prime - n_1 \vert >  n_1,  \; N \to \infty.
$

Thus perturbations induced by the ${\cal E}$  change crucially the statistics of $a=a_1$.   
On the other hand, the statistical deviation for $a=a_2$ is relatively small:

$\vert n_2^\prime - n_2 \vert \leq  n_2,  \; N \to \infty.
$

Thus perturbations induced by the ${\cal E}$  change slightly statistics of $a=a_2$.   
Orthogonality relation (\ref{OR}) implies that  phases 
$\theta_1$ and $\theta_2$ are not independent:
\begin{equation}
\label{ORH}
\cosh \;\theta_1 {\bf p}_1 +  \cos \; \theta_2 {\bf p}_2 =0 .
\end{equation}
Thus
\begin{equation}
\label{ORH1}
\frac{\cos\;\theta_2}{\cosh \; \theta_1}= - \frac{{\bf p}_1}{{\bf p}_2} .
\end{equation}
We get the {\it hyperbolic/trigonometric} probabilistic rule:
\begin{equation}
\label{H1}
{\bf p}_1^\prime = {\bf p}_1 (1 + \cosh \; \theta_1)= 2 {\bf p}_1 \cosh^2 \frac{\theta_1}{2} \;,
\end{equation}
\begin{equation}
\label{H2}
{\bf p}_2^\prime = {\bf p}_2 (1 + \cos\; \theta_2)= 2 {\bf p}_2 \cos^2 \frac{\theta_2}{2} \;.
\end{equation}

{\bf Example.} Let ${\bf p}_1 = 1/4, {\bf p}_2 = 3/4$ and let 
${\bf p}_1^\prime= a, {\bf p}_2^\prime= 1-a,$ where 
$a \in (1/2, 1].$ We cannot represent
${\bf p}_1^\prime= 2 {\bf p}_1 \cos^2 \theta$
for any $\theta.$ Here we need to use the representation:
${\bf p}_1^\prime= 2 {\bf p}_1 \cosh^2 \theta,$
where $\theta$  changes between 0 and $\rm{arcosh}\; \sqrt{2}$
for $a$ changing between 1/2 and 1. Thus there is no usual trigonometric wave. There is a kind
of hyperbolic wave.

{\bf Remark.} In fact, the use of the $\cos \theta$ (and $\cosh \theta)$ representations of
the coefficients $\lambda_i$ is motivated by the quantum formalism (and even the classical
field theory, compare with P. Dirac [8]). In principle we can represent $\vert \lambda\vert\leq 1$
as $\lambda= f(\theta)$ where $f$ is any function $\vert f \vert \leq 1.$ Dependence
$f(\theta)$ is related to the dependence of the preparation procedure on some parameter
${\cal E}={\cal E}(\theta).$ In many experiments $f$ is determined by the space geometry
of the experiment. For example, in the experiments with the light polarization
we use the Euclidean geometry to modify the preparation procedure ${\cal E}={\cal E}(\theta).$
This induces the $\cos$-factor. It is possible to construct families of preparation procedures 
${\cal E}(\theta)$ connected with other functions $f(\theta).$

\section{Experiments of `hyperbolic polarization'}

We note that both types of probablistic  transformations (purely trigonometric and
hyper-trigonometric) do happen in quantum experiments and
are routinely observed. If we have prepared an ensemble of
dichotomic systems such that they exhibit property $a_1$ with probability
${\bf p}_1$ and property $a_2$ with probability ${\bf p}_2$ (with ${\bf p}_1+ {\bf p}_2=1),$ 
then you can always change parameters of the preparation procedure in such a way
that input probablities ${\bf p}_1, {\bf p}_2$ will be transformed into output
probablities ${\bf p}_1^\prime, {\bf p}_2^\prime,$ where ${\bf p}_1^\prime, {\bf p}_2^\prime$
can have ANY
values between 0 and 1 (and of course, we again have ${\bf p}_1^\prime+{\bf p}_2^\prime=1).$

In experiments with polarized neutrons such experiments are often done.
We can prepare an arbitrary spin state, whose projection in a Stern-Gerlach
apparatus gives you the two possible outcomes with probabilities ${\bf p}_1, {\bf p}_2.$
But before the projection you can let the spin pass through a magnetic field
around which it precesses, and then you can adjust ANY DESIRED ${\bf p}_1^\prime, {\bf p}_2^\prime.$

\section{Physical consequences}

1). We demonstrated that there is nothing mysterious in so called
the `quantum probabilistic rule', compare with P. Dirac [8], R. Feynman [9]
(see also [10]). This rule can be derived by taking into account perturbation
effects of preparation/measurement procedures.\footnote{In fact, P. Dirac pointed out that
such effects play the crucial role in the creation of quantum behaviour [8].
However, he did not pay attention to the possibility to find
the purely probabilistic explanation of the origin of 
`quantum probabilistic rule.' He must use wave arguments [8]: ``If the two components are now made to interfere,
we should require a photon in one component to be able to interfere with one in the other."
The same was done by N. Bohr [5] who had to introduce the principle of complementarity 
to combine corpuscular and wave properties of elementary particles.}

2). We need not apply to `wave features' of quantum systems. The model 
can be purely corpuscular. It seems to be that quantum waves are just
probablistic waves (induced by statistical perturbations of preparation 
procedures).

3). The transition from classical behaviour to quantum behaviour is the 
transition from experiments with negligibly small statistical deviations
to experiments with nontrivial statistical deviations, compare 
with P. Dirac [8].

4). Our analysis did not demonstrate any difference in probabilistic
behaviour of macro and micro systems. `Quantum probabilistic behaviour' 
might be observed in experiments with macro systems, `polarization'
and `interference' of macro balls.

Our investigation was induced by investigations of J. Summhammer [11] 
on the origin of the quantum probablistic rule. I would like to thank
J. Summhammer for numerous (critical) discussions.

{\bf References}

[1] G. Ludwig, {\it Foundations of Quantum Mechanics.} (Springer Verlag, Berlin, 1983);
L.E. Ballentine, {\it Quantum Mechanics.} (Englewood Cliffs, New Jersey, 1989);
A. Peres, {\it Quantum Theory: Concepts and Methods.} (Kluwer Academic Publishers, 1994);
P. Busch, M. Grabowski, P.J. Lahti, {\it Operational Quantum Physics.}
(Springer Verlag, 1995).

[2] W. De Muynck, W. De Baere, H. Martens,
Found. of Physics, {\bf 24}, 1589--1663 (1994);
I. Pitowsky,  Phys. Rev. Lett, {\bf 48}, N.10, 1299-1302 (1982);
S. P. Gudder,  J. Math Phys., {\bf 25}, 2397- 2401 (1984);
W. De Muynck, J.T. Stekelenborg,  Annalen der Physik, {\bf 45},
N.7, 222-234 (1988); L. Accardi, {\it Urne e Camaleoni: Dialogo sulla realta,
le leggi del caso e la teoria quantistica.} (Il Saggiatore, Rome, 1997);
A. Khrennikov, J. Math. Phys., {\bf 41}, 1768-1777 (2000).

[3] N. Bohr, Phys. Rev., {\bf 48}, 696 (1935).

[4] A. Khrennikov, {\it Ensemble fluctuations and the origin of quantum probabilistic
rule.} Report MSI, Vaxjo University, N.90, October, (2000).

[5] R.  von Mises, {\it The mathematical theory of probability and
 statistics}. (Academic, London,  1964).
 
[6] A.Yu. Khrennikov,  {\it Interpretations of 
probability.} (VSP Int. Publ., Utrecht, 1999).
 
[7]  A. N. Kolmogoroff, {\it Grundbegriffe der Wahrscheinlichkeitsrechnung.}
(Springer Verlag, Berlin, 1933); reprinted:
{\it Foundations of the Probability Theory}. 
(Chelsea Publ. Comp., New York, 1956).

[8] P. A. M.  Dirac, {\it The Principles of Quantum Mechanics.}
(Claredon Press, Oxford, 1995).

[9] R. Feynman and A. Hibbs, {\it Quantum Mechanics and Path Integrals.}
(New York, 1965).

[10] B. d'Espagnat, {\it Veiled Reality. An anlysis of present-day
quantum mechanical concepts.} (Addison-Wesley, 1995). 

[11] J. Summhammer, Int. J. Theor. Physics, {\bf 33}, 171-178 (1994);
Found. Phys. Lett. {\bf 1}, 113 (1988); Phys.Lett., {\bf A136,} 183 (1989).

\end{document}